\documentclass[preprint]{aastex}

\usepackage{amsmath,amssymb,bm,mathrsfs}

\newcommand{\<}{{\kern-5pt}}

\newcommand{\thrj}[6]{\biggl(
	\begin{matrix}
	#1&#2&#3\\
	#4&#5&#6\\
	\end{matrix}\biggr)}

\newcommand{\apx}[1]{^{\mbox{\tiny{(#1)}}}}

\begin{document}

\title{Explicit Form of the Radiative and Collisional Branching Ratios\break
in Polarized Radiation Transport with Coherent Scattering}
\author{R.\ Casini,$^a$ T.\ del Pino Alem\'an,$^a$ and R.\ Manso Sainz$^b$}

\affil{$^a$High Altitude Observatory, National Center for Atmospheric
Research,\footnote{The National Center for Atmospheric Research is sponsored
by the National Science Foundation.}\break
P.O.~Box 3000, Boulder, CO 80307-3000, U.S.A.}
\affil{$^b$Max-Planck-Institut f\"ur Sonnensystemforschung,\break
Justus-von-Liebig-Weg 3, 37077 G\"ottingen, Germany}

\begin{abstract}
We consider the vector emissivity of the polarized radiation transfer 
in a $\Lambda$-type atomic transition, which we recently proposed in 
order to account for both CRD and PRD contributions to the
scattered radiation. 
This expression can concisely be written as
\begin{displaymath}
\bm{\varepsilon}=\left(\bm{\varepsilon}\apx{1}
		      -\bm{\varepsilon}\apx{2}_{\rm f.s.} \right)
	+\bm{\varepsilon}\apx{2}\;,
\end{displaymath}
where $\bm{\varepsilon}\apx{1}$ and $\bm{\varepsilon}\apx{2}$ are
the emissivity terms describing, respectively, one-photon and 
two-photon processes in a $\Lambda$-type atom,
and where ``f.s.'' means that the corresponding term
must be evaluated assuming an appropriate ``flat spectrum'' average
of the incident radiation across the spectral line.
In this follow up study, we explicitly consider the expressions of 
these various terms for the case of a polarized multi-term atom, 
in order to derive the algebraic forms of the branching ratios 
between the CRD and PRD contributions to the emissivity.
In the limit of a two-term atom with non-coherent lower-term, 
our results are shown to be in full agreement with those recently 
derived by \cite{Bo17}.
\end{abstract}

\section{Introduction}

In recent years, there has been a renewed interest of the solar
physics community in the theoretical problem of the generation and
transport of polarized radiation in the presence of frequency 
redistribution mechanisms 
(partial redistribution, or PRD). This is motivated by the importance of 
modeling the polarization patterns of deep chromospheric lines, for the
purpose of inferring the vector magnetic fields in the upper   
layers of the solar atmosphere. Examples of important diagnostic lines,
whose spectro-polarimetric patterns require to account for PRD effects
in order to be understood,
are the first few lines of the Lyman and Balmer series of hydrogen, the 
sodium D$_1$--D$_2$ doublet at 589.3\,nm, the h and k lines of singly ionized 
magnesium at 280\,nm, and the complex system
of singly ionized calcium encompassing the H and K lines around 395\,nm
and the infrared (IR) triplet system in the 850--866\,nm spectral region.

Over the past two decades, advances in the theoretical description of 
the partial frequency redistribution of polarized radiation in
astrophysical plasmas \citep{La97,Bo97a,Bo97b,BS99,Ca14,Bo16,CM16} 
have allowed for the first time the numerical modeling of some of these 
important spectral diagnostics of the solar chromosphere. This has 
yielded the first numerical predictions of the polarimetric 
signature of magnetic fields on the emergent Stokes profiles of 
spectral lines that are formed under PRD conditions, for 
magnetic field strengths that extend down to 
the regime of the Hanle effect, and hence adequate for the investigation 
of the magnetism of the quiet Sun chromosphere \citep{TB11,So14,dPA16,Al16}

\begin{figure}[t!]
\centering
\includegraphics[height=.34\vsize]{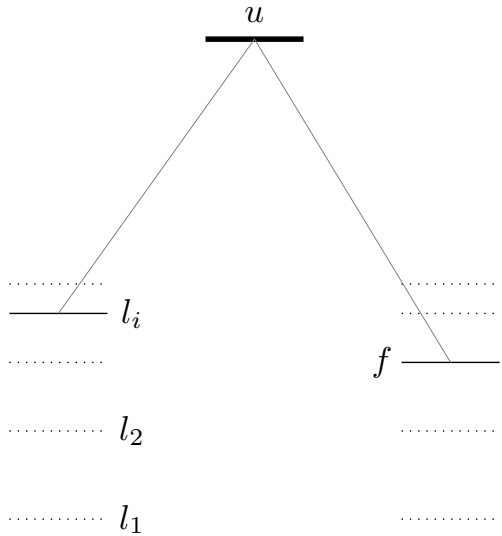}
\caption{Schematic diagram of the $\Lambda$-type multi-term atom.
In order to correctly describe the polarization properties of the outgoing light
in the $u\to f$ transition, all lower terms $l_i$ that are radiatively
or collisionally connected to the upper term $u$, including the final 
term $f$, must be taken into account.
\label{fig:atom_model}}
\end{figure}

Among the various theoretical approaches to the problem of polarized radiative 
transfer (RT) with PRD, the two advanced by \cite{Bo97a,Bo97b,Bo16} and by Casini 
and collaborators \citep{Ca14,CM16} have in particular come under some 
in-depth scrutiny. Both
approaches rely on a derivation of the respective formalisms from the 
first principles of non-relativistic quantum electrodynamics, 
yet they have led to seemingly different interpretations of the driving
physical processes of radiation scattering 
\cite[see the discussions in][]{Ca14,Bo16,Bo17}. 

The problem of determining the proper expressions of the \emph{branching 
ratios} for the fully redistributed (CRD) and the partially 
redistributed contributions of the polarized emissivity to the 
RT equation in a collisional plasma has recently been 
considered within both approaches. This is one of the key
problems for the 
validation of any theoretical description of radiation scattering, and 
ultimately for its applicability to the numerical modeling of polarized 
RT in a magnetized chromosphere. \cite{Ca17} have derived 
a very general form of the scattering emissivity, 
which is broadly applicable to numerical problems of polarized radiation 
scattering in multi-term atoms of the $\Lambda$-type \cite[][see also
Figure~\ref{fig:atom_model}]{CM16}. 
\cite{Bo17} has explicitly derived the expressions of the branching ratios 
for the CRD and PRD contributions to the emissivity redistribution
function for the case of a two-term atom with non-coherent and 
unpolarized lower term.

In this work we show how the two approaches lead to branching ratios
that are in full agreement, when the generalized form of the emissivity
of \cite{Ca17} is explicitly applied to the same atomic model considered by
\cite{Bo17}.

\section{General form of the polarized scattering emissivity}

We consider the general expression of the polarized radiation emissivity 
proposed by \cite{Ca17}, describing the partially coherent 
scattering of radiation in spectral lines from permitted 
transitions that are formed under PRD conditions. This
emissivity represents the source term of the vector transfer 
equation for polarized radiation,
\begin{equation} \label{eq:RT.gen}
\frac{d}{ds}\,\bm{S}(\omega_{k'},\bm{\hat k}') 
= -{\bf K}(\omega_{k'},\bm{\hat k}')\,
	\bm{S}(\omega_{k'},\bm{\hat k}')
+\bm{\varepsilon}(\omega_{k'},\bm{\hat k}')\;,
\end{equation}
where $\bm{S}\equiv(S_0,S_1,S_2,S_3)$ is the Stokes vector of the
polarized radiation field (of frequency $\omega_{k'}$ and propagation
direction $\bm{\hat k}'$), $\bf K$ is the 4$\times$4 polarized 
absorption matrix (responsible for both isotropic and dichroic 
absorption, as well as magneto-optical effects), and
\begin{equation} \label{eq:new.emiss}
\bm{\varepsilon}(\omega_{k'},\bm{\hat k}')
	\equiv \Bigl[\bm{\varepsilon}\apx{1}(\omega_{k'},\bm{\hat k}')
	-\bm{\varepsilon}\apx{2}(\omega_{k'},\bm{\hat k}')_{\rm f.s.}
	\Bigr]
+\bm{\varepsilon}\apx{2}(\omega_{k'},\bm{\hat k}')\;,
\end{equation}
is the generalized emissivity.
For a multi-term atomic system of the $\Lambda$-type (see 
Figure~\ref{fig:atom_model}), the first-order emissivity 
\emph{in the atomic frame of reference} is given by
\begin{eqnarray} \label{eq:incoherent.flat}
\varepsilon\apx{1}_i(\omega_{k'},\bm{\hat k}')
&=& \frac{1}{2\pi^2\sqrt3}\,\frac{e_0^2}{c^3}\,
	{\cal N}\omega_{k'}^4
	\sum_{uu'}\rho_{uu'}\sum_{f}\sum_{pp'}(-1)^{p'+1}\,
	(r_p)_{u'f}(r_{p'})_{uf}^* \\
&&{}\times
	\sum_{K'Q'}\sqrt{2K'+1}\,\thrj{1}{1}{K'}{-p}{p'}{-Q'}\,
	T^{K'}_{Q'}(i,\bm{\hat k}')
	\Bigl(
	\Phi_{fu'}^{+k'} + \bar\Phi_{fu}^{+k'}
	\Bigr)\;,\qquad (i=0,1,2,3) \nonumber
\end{eqnarray}
where we introduced the usual complex line profiles,
\begin{displaymath} \label{eq:Phi}
\Phi_{ab}^{\pm k} 
\equiv \frac{\rm i}{\omega_{ba}\mp\omega_k+{\rm i}\epsilon_{ba}}\;,
\end{displaymath}
and we indicated with $\bar\Phi$ the operation of complex conjugation.
For the definition of all the other physical quantities in 
equation~(\ref{eq:incoherent.flat})
and in the following equations, we refer to \cite{Ca14}.
Making use of the conjugation properties of those various quantities, it is
possible to show that this emissivity term is purely real. In
particular, this is a direct consequence of the hermiticity of the
density-matrix $\rho_{uu'}$ for the upper term.

The contribution (\ref{eq:incoherent.flat}) to the general emissivity is associated with 
excitation mechanisms of the upper term that lead to the production of completely 
redistributed radiation, i.e., to fully non-coherent scattering of the incident 
radiation. As
an example, this term accounts for the emission of radiation via the spontaneous 
decay of a collisionally excited level.

For typical applications to astrophysical plasmas, it is safe to assume
that the incident radiation field is highly diluted (\emph{weak radiation
field approximation}), so that the radiative lifetime of the lower
atomic levels is practically infinite compared to their collisional 
lifetime. 
Under this assumption, radiation absorption does not contribute
appreciably to the population of the upper state \citep{Ca14}. 
Therefore, that population is for practical purposes completely determined by 
the balance between collisional excitation and spontaneous de-excitation 
(because of the weak radiation
field assumption, all stimulated radiation processes can also be neglected).

The second-order emissivity, which accounts for two-photon processes 
in a $\Lambda$-type atom leading to partially redistributed 
scattering, in the atomic 
frame of reference is given by \citep[see][]{Ca14}
\begin{eqnarray} \label{eq:2emiss.1}
\varepsilon\apx{2}_i(\omega_{k'},\bm{\hat k}')
&=&\frac{4}{3}\frac{e_0^4}{\hbar^2 c^4}\,{\cal N}\omega_{k'}^4
	\sum_{ll'}\rho_{ll'}\sum_{uu'f}
	\sum_{qq'}\sum_{pp'}(-1)^{q'+p'}\,
	(r_q)_{ul}(r_{q'})^\ast_{u'l'}
	(r_p)_{u'f}(r_{p'})^\ast_{uf} \nonumber \\
&&\times
	\sum_{KQ}\sum_{K'Q'}\sqrt{(2K+1)(2K'+1)}\,
	\thrj{1}{1}{K}{-q}{q'}{-Q}
	\thrj{1}{1}{K'}{-p}{p'}{-Q'}\,
	T^{K'}_{Q'}(i,\bm{\hat k}') \nonumber \\
&&\times \int_0^\infty d\omega_k\,
	\Bigl(
	\Psi_{u'l',ful}^{-k,+k'-k} + \bar\Psi_{ul,fu'l'}^{-k,+k'-k}
	\Bigr)\,
	J^K_Q(\omega_k)\;,\qquad (i=0,1,2,3)
\end{eqnarray}
where
\begin{eqnarray}
\Psi_{ab,cde}^{\pm h,\pm k\pm l}
&\equiv&
\frac{-{\rm i}}{
(\omega_{ac}\pm\omega_h\mp\omega_l\mp\omega_k
	+{\rm i}\epsilon_a+{\rm i}\epsilon_c)
(\omega_{ad}\pm\omega_h\mp\omega_l+{\rm i}\epsilon_a+{\rm i}\epsilon_d)
(\omega_{ae}\pm\omega_h+{\rm i}\epsilon_a+{\rm i}\epsilon_e)
} \nonumber \\
&+&\frac{-{\rm i}}{
(\omega_{ac}\pm\omega_h\mp\omega_l\mp\omega_k
	+{\rm i}\epsilon_a+{\rm i}\epsilon_c)
(\omega_{bc}\mp\omega_l\mp\omega_k+{\rm i}\epsilon_b+{\rm i}\epsilon_c)
(\omega_{cd}\pm\omega_k-{\rm i}\epsilon_c+{\rm i}\epsilon_d)
} \nonumber \\
&-&\frac{-{\rm i}}{
(\omega_{ad}\pm\omega_h\mp\omega_l+{\rm i}\epsilon_a+{\rm i}\epsilon_d)
(\omega_{bd}\mp\omega_l+{\rm i}\epsilon_b+{\rm i}\epsilon_d)
(\omega_{cd}\pm\omega_k-{\rm i}\epsilon_c+{\rm i}\epsilon_d)
}\;,
\end{eqnarray}
and with $\bar\Psi$ we indicated the complex conjugate of the profile
$\Psi$.
The frequency integral in equation~(\ref{eq:2emiss.1}) can be rewritten 
in terms of the redistribution function ${\cal R}(\omega_k,\omega_{k'})$ 
for a three-term $\Lambda$-atom, using the definition (4) of 
\citeauthor{Ca17} (\citeyear{Ca17};
see also \citealt{Ca14}, equation~(6)),
\begin{equation} \label{eq:red.gen}
{\cal R}(\Omega_u,\Omega_{u'};
	\Omega_l,\Omega_{l'},\Omega_{f};
	\omega_k,\omega_{k'})
= {\rm i}(\Omega_u-\Omega_{u'}^\ast)
	\Bigl(
	\Psi_{u'l',ful}^{-k,+k'-k} + \bar\Psi_{ul,fu'l'}^{-k,+k'-k}
	\Bigr)\;,
\end{equation}
where we introduced the complex atomic frequencies
$\Omega_a=\omega_a-{\rm i}\epsilon_a$.

%

The subscript ``f.s.'' in one of the two second-order terms of the 
general emissivity in equation~(\ref{eq:new.emiss}) means that such term must 
be evaluated \emph{as if} the incident radiation field were spectrally flat,
i.e., described by some average $J^K_Q(\bar\omega$) of the incident radiation 
field tensor across the atomic transition. 
For the moment, we leave aside the question of what such a suitable average 
should look like, and simply observe that an immediate result of that 
averaging procedure is the ability to perform the integration over the incident 
frequency $\omega_k$ in equation~(\ref{eq:2emiss.1}).
This is simply attained by recalling the following integral norm of the
redistribution function \cite[cf.][equation (15)]{Ca14},
\begin{equation} \label{eq:integral2}
\int_{0}^\infty d\omega_{k}\,
	\Bigl(
	\Psi_{u'l',ful}^{-k,+k'-k} + 
	\bar\Psi_{ul,fu'l'}^{-k,+k'-k}
	\Bigr)
=\frac{2\pi}{\epsilon_{uu'}+{\rm i}\omega_{uu'}}\,
	\Bigl(
	\Phi_{fu'}^{+k'} + \bar\Phi_{fu}^{+k'}
	\Bigr)\;.
\end{equation}
Substitution of this integral expression into
equation~(\ref{eq:2emiss.1}) immediately yields
\begin{eqnarray} \label{eq:coherent.flat}
\varepsilon\apx{2}_i(\omega_{k'},\bm{\hat k}')_{\rm f.s.}
&=&
	\frac{4}{3}\frac{e_0^4}{\hbar^2 c^4}\,{\cal N}\omega_{k'}^4
	\sum_{ll'}\rho_{ll'}\sum_{uu'f}
	\sum_{qq'}\sum_{pp'}(-1)^{q'+p'}\,
	(r_q)_{ul}(r_{q'})^\ast_{u'l'}
	(r_p)_{u'f}(r_{p'})^\ast_{uf} \nonumber \\
&&{}\times
	\sum_{KQ}\sum_{K'Q'}\sqrt{(2K+1)(2K'+1)}\,
	\thrj{1}{1}{K}{-q}{q'}{-Q}
	\thrj{1}{1}{K'}{-p}{p'}{-Q'}\,
	T^{K'}_{Q'}(i,\bm{\hat k}') \nonumber \\
&&\kern 2cm{}\times
	\frac{2\pi}{\epsilon_{uu'}+{\rm i}\omega_{uu'}}\,
	\Bigl(
	\Phi_{fu'}^{+k'} + \bar\Phi_{fu}^{+k'}
	\Bigr)\,
	J^K_Q(\bar\omega) \nonumber \\
&=& 
	\frac{1}{2\pi^2\sqrt3}\,\frac{e_0^2}{c^3}\,
	{\cal N}\omega_{k'}^4
	\sum_{uu'} \Biggl[
	\frac{16\pi^3}{\sqrt3}\frac{e_0^2}{\hbar^2 c}\,
	\frac{1}{\epsilon_{uu'}+{\rm i}\omega_{uu'}}
	\sum_{ll'}\rho_{ll'}
	\sum_{qq'}(-1)^{q'+1}\,
	(r_q)_{ul}(r_{q'})^\ast_{u'l'} \nonumber \\
&&\kern 3.5cm{}\times
	\sum_{KQ}\sqrt{2K+1}\,
	\thrj{1}{1}{K}{-q}{q'}{-Q}
	J^K_Q(\bar\omega)
	\Biggr] \nonumber \\
\noalign{\allowbreak}
&&{}\times
	\sum_{f}\sum_{pp'}(-1)^{p'+1}\,
	(r_p)_{u'f}(r_{p'})^\ast_{uf}
	\sum_{K'Q'}\sqrt{2K'+1}\,
	\thrj{1}{1}{K'}{-p}{p'}{-Q'} \nonumber \\
&&\kern 3.5cm{}\times
	T^{K'}_{Q'}(i,\bm{\hat k}')
	\Bigl(
	\Phi_{fu'}^{+k'} + \bar\Phi_{fu}^{+k'}
	\Bigr)\;.\qquad (i=0,1,2,3)
\end{eqnarray}

A direct comparison of this result with equation~(\ref{eq:incoherent.flat}) 
shows that the two expressions of $\varepsilon\apx{1}$ and 
$\varepsilon\apx{2}_{\rm f.s.}$ coincide when the quantity 
inside the square brackets in equation~(\ref{eq:coherent.flat}) can be
identified with the upper-term atomic density matrix $\rho_{uu'}$
\cite[][Section 7]{Ca14}.
This is indeed the case for the solution of the \emph{first-order} statistical 
equilibrium (SE) problem for the multi-term atom of the $\Lambda$-type 
\emph{in the presence of only radiative processes}, involving one-photon
absorption and emission, when stimulated emission is negligible (weak radiation 
field approximation).
In this limit, the identification of $\rho_{uu'}$ with the expression
inside the square brackets in equation~(\ref{eq:coherent.flat}) implies that 
$J^K_Q(\bar\omega)$ must correspond to the integral average of the 
incident radiation field tensor that appears in the transfer rate 
for radiative absorption of the first-order SE problem
\cite[cf.][]{La83,La84}. This rate is given by
\begin{eqnarray} \label{eq:tA}
t_{\rm A}(uu',ll')
&=& \frac{16\pi^3}{\sqrt3}\frac{e_0^2}{\hbar^2 c}\,
	\sum_{qq'} (-1)^{q+q'} (r_q)_{ul}
	(r_{q'})_{u'l'}^\ast
	\sum_{KQ}\sqrt{2K+1}\,
	\thrj{1}{1}{K}{-q}{q'}{-Q} \nonumber \\
&&{}\times \frac{1}{2\pi}\int_0^\infty d\omega_k\,
	\Bigl( \Phi_{l'u'}^{+k} + \bar\Phi_{lu}^{+k} \Bigr)\,
	J^K_Q(\omega_k)\;,
\end{eqnarray}
hence,
\begin{equation} \label{eq:Javg.def}
J^K_Q(\bar\omega)=
J^K_Q(\omega_{ul;u'l'})\equiv
\frac{1}{2\pi}
\int_{0}^\infty d\omega_{k}\,
	\Bigl(
	\Phi_{l'u'}^{+k} + \bar\Phi_{lu}^{+k}
	\Bigr)\,J^K_Q(\omega_k)\;.
\end{equation}
For the following development, we adopt equation~(\ref{eq:Javg.def}) 
as the proper definition of the integral average $J^K_Q(\bar\omega)$ 
entering equation~(\ref{eq:coherent.flat}).

It must be noted that such ``average'' generally takes different values 
for different pairs of atomic transition components $(ul;u'l')$. In
this sense, the adoption of equation~(10) actually implies a full 
departure of the SE solution from the limitations of the flat-spectrum
approximation, which instead requires that the incident radiation field 
tensor $J^K_Q(\omega_k)$ must be structureless over the entire spectral 
range spanned by \emph{all} the components of the atomic transition. 
This is a critical aspect of our formalism, since the flat-spectrum 
approximation is instead a necessary condition for the physical
consistency of the first-order RT problem, where the
emissivity term is represented exclusively by $\bm{\varepsilon}\apx{1}$.
In our case, instead, use of the expression (\ref{eq:new.emiss}) for the
total emissivity allows us to relax the condition of the flat-spectrum
approximation in the first-order SE problem without breaking the
internal consistency of the description of the scattering process.

In contrast, the appearance of the average radiation tensor 
$J^K_Q(\bar\omega)$ in the $\bm{\varepsilon}\apx{2}_{\rm f.s.}$
emissivity does require replacing the incident radiation field in 
$\bm{\varepsilon}\apx{2}$ by an appropriate constant average 
\emph{for each} 
of the transition components, according to the definition 
(\ref{eq:Javg.def}). Hence the reason to dub that emissivity 
term as a ``flat spectrum'' contribution to the total emissivity.

In many practical cases of interest to solar 
polarimetry, the incident radiation field will largely be constant over 
the typical frequency separation of the Zeeman components of the line
in the presence of a magnetic field, whereas it might have a significant 
spectral modulation over the frequency range of the atomic fine 
structure, depending on the nature of the multiplets. This is certainly 
the case, e.g., for well-known doublets of the solar spectrum, such as 
\ion{Mg}{2} h--k, \ion{Ca}{2} H--K, or \ion{Na}{1} D$_1$--D$_2$, 
whereas for close spectral multiplets, such as \ion{H}{1} Ly$_\alpha$ 
and H$_\alpha$, the spectral modulation of the incident radiation field 
across the fine structure is generally unimportant.\footnote{This last 
assumption may become invalid in the presence of plasma velocity 
gradients in the line formation region, if the incident radiation 
is significantly Doppler shifted with respect to the atomic rest 
wavelength of the line.}
Therefore, in many applications it might be possible to
consider only a reduced set of integral averages of the type
$J^K_Q(\omega_{J_uJ_l;J_u'J_l'})$, thus simplifying the
numerical implementation of the line formation problem.

It is important to note that the exact cancellation between 
$\bm{\varepsilon}\apx{1}$ and
$\bm{\varepsilon}\apx{2}_{\rm f.s.}$ \emph{in the purely radiative case}
embodies our original assumption of infinite radiative lifetime of 
the lower terms, according to which no population of the upper term can 
actually be created through radiative absorption. In this second-order
framework, the two independent, first-order processes of absorption and 
re-emission of a single photon are instead replaced by a single, coherent
two-photon process represented by $\bm{\varepsilon}\apx{2}$.

In order to combine the contributions (\ref{eq:incoherent.flat}),
(\ref{eq:2emiss.1}), and (\ref{eq:coherent.flat}) into the
general emissivity (\ref{eq:new.emiss}), in the presence of
\emph{both} radiative and collisional processes, we must formally solve
the SE problem of the atomic system for the density matrix $\rho_{uu'}$ 
of the upper state under such more general case.

For an isotropic 
distribution of colliding perturbers, assuming the validity of the impact
approximation for collisions, it can be shown \cite[see, e.g.,][]{Be13} that 
the algebraic solution for $\rho_{uu'}$ can be generalized beyond the purely 
radiative expression inside the square brackets of 
equation~(\ref{eq:coherent.flat}).
Simply put, the radiative width $\epsilon_{uu'}$ for the upper term is augmented 
by the inverse lifetime for collisional relaxation of the excited levels, 
whereas a \emph{collisional} excitation term is added to the SE problem
alongside the \emph{radiative} excitation term corresponding to the $t_{\rm A}$ 
transfer rate (\ref{eq:tA}). The 
dominant effect of this last contribution from collisional excitation is
the tendency of the system towards thermalization of the atomic populations, 
with a corresponding contribution to the first-order emissivity that essentially 
corresponds to a Planckian radiation at the equilibrium temperature of the plasma.
Accordingly, the solution density matrix for the upper term appearing in
equation~(\ref{eq:incoherent.flat}) can be written as the sum of a
radiative and a collisional part \citep{Ca17},
\begin{displaymath}
\rho_{uu'}=\rho_{uu'}\apx{rad}+\rho_{uu'}\apx{coll}\;,
\end{displaymath}
and correspondingly we can write, from equation~(\ref{eq:incoherent.flat}),
\begin{equation} \label{eq:split.emiss}
\bm{\varepsilon}\apx{1}(\omega_{k'},\bm{\hat k}')=
\bm{\varepsilon}\apx{1;rad}(\omega_{k'},\bm{\hat k}')
+\bm{\varepsilon}\apx{1;coll}(\omega_{k'},\bm{\hat k}')\;.
\end{equation}

In the absence of elastic collisions, 
the (nearly Planckian) term in 
equation~(\ref{eq:split.emiss}) associated with $\rho_{uu'}\apx{coll}$ 
is the only completely redistributed contribution to the scattered radiation, 
since the contribution from $\rho_{uu'}\apx{rad}$ is exactly compensated
by $\bm{\varepsilon}\apx{2}_{\rm f.s.}$ (cf.\ 
equation~(\ref{eq:coherent.flat}), and the discussion after
equation~(\ref{eq:Javg.def})).
When instead elastic collisions are present, the denominator
${\rm i}(\Omega_u-\Omega_{u'}^\ast)$ that 
appears in the expression of $\bm{\varepsilon}\apx{2}$ (cf.\
equation~(\ref{eq:red.gen})) must also account for the
additional level width $\Gamma_{uu'}$ due to the perturbation of the atomic 
levels by the elastic colliders.
In contrast, because elastic collisions do not affect the population 
balance in the first-order SE problem, a corresponding contribution is 
instead missing from the denominator $\epsilon_{uu'}+{\rm i}\omega_{uu'}$ 
appearing in the formal expression of
$\rho_{uu'}\apx{rad}$. Hence, an exact compensation between
$\bm{\varepsilon}\apx{1;rad}$ and $\bm{\varepsilon}\apx{2}_{\rm f.s.}$ generally 
no longer occurs \emph{in the presence of elastic collisions}. 

In conclusion, we can rewrite the general emissivity (\ref{eq:new.emiss}) as
\begin{eqnarray} \label{eq:new.emiss.1}
\bm{\varepsilon}(\omega_{k'},\bm{\hat k}')
&=&
\bm{\varepsilon}\apx{1;coll}(\omega_{k'},\bm{\hat k}')+
\bm{\tilde\varepsilon}(\omega_{k'},\bm{\hat k}') \nonumber \\
&=&
\bm{\varepsilon}\apx{1;coll}(\omega_{k'},\bm{\hat k}')+
\left[\bm{\varepsilon}\apx{1;rad}(\omega_{k'},\bm{\hat k}')-
\bm{\varepsilon}\apx{2}_{\rm f.s.}(\omega_{k'},\bm{\hat k}')\right]
+\bm{\varepsilon}\apx{2}(\omega_{k'},\bm{\hat k}')\;,
\end{eqnarray}
and recalling equations~(\ref{eq:incoherent.flat}),
(\ref{eq:2emiss.1}), and (\ref{eq:coherent.flat}), after some
straightforward algebra, we find
\begin{eqnarray} \label{eq:tilde.emiss}
\tilde\varepsilon_i(\omega_{k'},\bm{\hat k}')
&=&\frac{4}{3}\frac{e_0^4}{\hbar^2 c^4}\,{\cal N}\omega_{k'}^4
	\sum_{ll'}\rho_{ll'}\sum_{uu'f}
	\sum_{qq'}\sum_{pp'}(-1)^{q'+p'}\,
	(r_q)_{ul}(r_{q'})^\ast_{u'l'}
	(r_p)_{u'f}(r_{p'})^\ast_{uf} \nonumber \\
&&\times
	\sum_{KQ}\sum_{K'Q'}\sqrt{(2K+1)(2K'+1)}\,
	\thrj{1}{1}{K}{-q}{q'}{-Q}
	\thrj{1}{1}{K'}{-p}{p'}{-Q'}\,
	T^{K'}_{Q'}(i,\bm{\hat k}') \nonumber \\
&&\times \int_0^\infty d\omega_k\;
	\mathscr{R}(\Omega_u,\Omega_{u'};
	\Omega_l,\Omega_{l'},\Omega_{f};
	\omega_k,\omega_{k'})\,
	J^K_Q(\omega_k)\;,\qquad (i=0,1,2,3)
\end{eqnarray}
where we introduced the \emph{generalized} 
redistribution function in the atomic frame of reference,
\begin{eqnarray} \label{eq:gen.R}
{\mathscr{R}}(\Omega_u,\Omega_{u'};
	\Omega_l,\Omega_{l'},\Omega_{f};
	\omega_k,\omega_{k'})
&=& \frac{1}{\epsilon_{uu'}+\Gamma_{uu'}+{\rm i}\omega_{uu'}}\,
{\cal R}(\Omega_u,\Omega_{u'};
	\Omega_l,\Omega_{l'},\Omega_{f};
	\omega_k,\omega_{k'}) \\
&&\kern -2cm + \left( \frac{1}{\epsilon_{uu'}+{\rm i}\omega_{uu'}}
         - \frac{1}{\epsilon_{uu'}+\Gamma_{uu'}+{\rm i}\omega_{uu'}}
           \right)\!
	\Bigl(
	\Phi_{l'u'}^{+k} + \bar\Phi_{lu}^{+k}
	\Bigr)
	\Bigl(
	\Phi_{fu'}^{+k'} + \bar\Phi_{fu}^{+k'}
	\Bigr)\;, \nonumber
\end{eqnarray}
and where ${\cal R}(\omega_k,\omega_{k'})$ is given by 
equation~(\ref{eq:red.gen}), while we considered that 
${\rm i}(\Omega_u-\Omega_{u'}^\ast)
=\epsilon_{uu'}+\Gamma_{uu'}+{\rm i}\omega_{uu'}$, in the additional 
presence of elastic collisions. According to the previous discussion, the 
average radiation tensor (\ref{eq:Javg.def}) must be adopted for both 
equations~(\ref{eq:incoherent.flat}) and (\ref{eq:coherent.flat}) in
order to arrive at equation~(\ref{eq:gen.R}). 

The non-Planckian emissivity (\ref{eq:tilde.emiss}) can further be specialized 
to particular atomic structures, such as the multi-term atom of the
$\Lambda$-type in the $LS$-coupling scheme, with and without hyperfine 
structure, and the multi-level atom. This is immediately accomplished by
taking 
equations~(4), (7), and (8) of \cite{CM16}, respectively, for those
three atomic models, and substituting the redistribution integral of 
equation~(\ref{eq:tilde.emiss}) in those equations, using the definition
(\ref{eq:gen.R}), while at the same time omitting from those earlier
equations the factor $1/(\epsilon_{uu'}+{\rm i}\omega_{uu'})$, as this
is already accounted for, in a more general form, by the
redistribution function (\ref{eq:gen.R}).

With this straightforward exercise, we can verify that our second-order
emissivity (\ref{eq:tilde.emiss}) and generalized redistribution function 
(\ref{eq:gen.R}) agree with those derived by \cite{Bo17} for the case of 
the two-term atom with non-coherent lower term (i.e., 
$\rho_{ll'}=\delta_{ll'}\rho_{ll}$) and infinitely sharp
lower levels (i.e., $\epsilon_l,\epsilon_f\to 0$), once the
redistribution function ${\cal R}(\omega_k,\omega_{k'})$ is replaced by 
the appropriate expression for this case \cite[][equation~(10)]{Ca14}.
In particular, depolarizing collisions can also be included analogously to
\cite{Bo17}, taking into consideration the corresponding contributions
to the first-order SE problem. In such case, a new level width contribution 
is added to the denominator of $1/(\epsilon_{uu'}+{\rm i}\omega_{uu'})$ 
in the second line of equation~(\ref{eq:gen.R}), 
as a consequence of this modification for the 
atomic density matrix solution of the SE problem \cite[see][]{Bo17}.

\section{Conclusions}

We derived the explicit form of the branching ratios for the
contributions of completely redistributed radiation and of partially
coherent scattering to the generalized emissivity in the radiative
transfer equation for polarized radiation. The expression for this 
emissivity was proposed by \cite{Ca17} in the implicit form of
equation~(\ref{eq:new.emiss}). In this work, we considered the 
algebraic expressions of its various contributions, and 
demonstrated how they combine to give rise to a new form of the 
second-order emissivity for partially coherent scattering in a two-term
atom proposed by \cite{Ca14} (see also \citealt{CM16} for its extension
to the three-term atom of the $\Lambda$-type). In this emissivity, 
the usual partial redistribution function is replaced by a more general 
expression, equation~(\ref{eq:gen.R}), which also takes into account 
the contribution of fully redistributed radiation caused by 
collisional processes.


In the limit of the two-term atom with non-coherent lower
term, the generalized redistribution function (\ref{eq:gen.R}) 
proves to be identical to the one recently derived by \cite{Bo17} through 
her own ab-initio formulation of the polarized scattering problem. This
verification confirms that the two independent formalisms of 
\cite{Bo97a,Bo97b,Bo16,Bo17} and of \cite{Ca14}, \cite{CM16}, and \cite{Ca17}, 
though seemingly different, are in fact complementary, and lead to exactly 
the same predictions for the properties of the partially redistributed 
polarized radiation scattered by a two-term atom.
In particular, \emph{the adoption of equation~(\ref{eq:new.emiss}) for 
the total emissivity of the radiative transfer problem---and of
equation~(\ref{eq:gen.R}) for the CRD and PRD branching ratios---allows us to
completely relax the limitation of the flat-spectrum approximation in
the solution of the first-order statistical equilibrium problem}
\cite[see also][]{Bo17}. Accordingly, the statistical
equilibrium problem underlying the formation of spectral lines
under PRD conditions corresponds to the original formulation of 
\cite{La83,La84}, as originally pointed out by
\cite{Bo97a,Bo97b}.

As a final remark, it is important to point out that while the derivation 
of the generalized redistribution function in the explicit form (\ref{eq:gen.R}) 
relied specifically on the choice of a reference frame at rest with the
atom, the implicit form given by equation~(\ref{eq:new.emiss}) is
instead generally valid, and thus it is immediately applicable to
numerical problems of polarized radiative transfer in model atmospheres. 
Such approach was followed by \cite{dPA16} for the modeling of the 
\ion{Mg}{2} h--k doublet in a magnetized chromosphere including PRD effects.

\begin{acknowledgments}
We thank the anonymous referee for helpful comments. We thank 
V. Bommier (LEISA, France) for a careful reading of the manuscript, and
for several suggestions on the presentation of this work.

\end{acknowledgments}
%


\begin{thebibliography}

\bibitem[\protect\citeauthoryear{Alsina Ballester, Belluzzi, \&
Trujillo Bueno}{2016}]{Al16}
Alsina Ballester, E., Belluzzi, L., \& Trujillo Bueno, J.~2016,
\apjl, 831, 15

\bibitem[\protect\citeauthoryear{Belluzzi, Landi Degl'Innocenti, \&
Trujillo Bueno}{2013}]{Be13}
Belluzzi, L., Landi Degl'Innocenti, E., \& Trujillo Bueno, J.~2013,
\aap, 551, 84

\bibitem[\protect\citeauthoryear{Bommier}{1997a}]{Bo97a}
Bommier, V.~1997a, \aap, 328, 706

\bibitem[\protect\citeauthoryear{Bommier}{1997b}]{Bo97b}
Bommier, V.~1997b, \aap, 328, 726

\bibitem[\protect\citeauthoryear{Bommier}{2016}]{Bo16}
Bommier, V.~2016, \aap, 591A, 59

\bibitem[\protect\citeauthoryear{Bommier}{2017}]{Bo17}
Bommier, V.~2017, \aap\ (in press)

\bibitem[\protect\citeauthoryear{Bommier \& Stenflo}{1999}]{BS99}
Bommier, V., \& Stenflo, J.~O.~1999, \aap, 350, 327

\bibitem[\protect\citeauthoryear{Casini et al.}{2014}]{Ca14}
Casini, R., Landi Degl'Innocenti, M., Manso Sainz, R., Landi Degl'Innocenti, M., 
\& Landolfi, M.~2014, \apj, 791, 94

\bibitem[\protect\citeauthoryear{Casini \& Manso Sainz}{2016}]{CM16}
Casini, R., \& Manso Sainz, R.~2016, \apj, 824, 135

\bibitem[\protect\citeauthoryear{Casini, del Pino Alem\'an, \& Manso Sainz}{2017}]{Ca17}
Casini, R., del Pino Alem\'an, T., \& Manso Sainz, R.~2017, \apj, 835, 114

\bibitem[\protect\citeauthoryear{del Pino Alem\'an, Casini, \& Manso Sainz}{2016}]{dPA16}
del Pino Alem\'an, T., Casini, R., \& Manso Sainz, R.~2016, \apjl, 830, L24

\bibitem[\protect\citeauthoryear{Landi Degl'Innocenti}{1983}]{La83}
Landi Degl'Innocenti, E. 1983, \solphys, 85, 3

\bibitem[\protect\citeauthoryear{Landi Degl'Innocenti}{1984}]{La84}
Landi Degl'Innocenti, E. 1984, \solphys, 91, 1

\bibitem[\protect\citeauthoryear{Landi Degl'Innocenti \& Landolfi}{2004}]{LL04}
Landi Degl'Innocenti, E., \& Landolfi, M. 2004, Polarization in
Spectral Lines (Dordrecht: Springer)

\bibitem[\protect\citeauthoryear{Landi Degl'Innocenti, Landi
Degl'{\allowbreak}Innocenti, \& Landolfi}{1997}]{La97}
Landi Degl'Innocenti, E., Landi Degl'Innocenti, M., \& Landolfi, M.\ 1997, 
in Proc.~Forum TH\'EMIS, Science with TH\'EMIS, ed.~N.~Mein \& 
S.~Sahal-Br\'echot (Paris: Obs.~Paris-Meudon), 59

\bibitem[\protect\citeauthoryear{Sowmya et al.}{2014}]{So14}
Sowmya, K., Nagendra, K. N., Sampoorna, M., \& Stenflo, J. O.~2014,
\apj, 793, 71

\bibitem[\protect\citeauthoryear{Trujillo Bueno, \v{S}t\v{e}p\'an, \&
Casini}{2011}]{TB11}
Trujillo Bueno, J., \v{S}t\v{e}p\'an, J., \& Casini, R.~2011, \apjl,
738, 11

\end{thebibliography}
\end{document}